# UNIFORM PLASMA OSCILLATIONS IN ELLIPSOID OF CONDUCTIVE MATERIAL

*UDC 533.92*

## Yuri Kornyushin

Maître Jean Brunschvig Research Unit, Chalet Shalva, Randogne, CH-3975

**Abstract**. *The influence of the shape of a sample on the type of uniform dipole collective electrons oscillations is discussed. In samples of a bulk shape uniform bulk dipole oscillation (Langmuir oscillation) cannot exist. It exists in samples of a thin slab shape only. As uniform bulk dipole oscillations cannot penetrate ellipsoidal samples of conductive material, they exist in the surface layer of a sample only (Mie oscillations). Frequencies of Mie oscillations are calculated for a sample of the shape of an arbitrary ellipsoid.*

### 1. INTRODUCTION

Collective oscillations are the most prominent features of the excitation spectrum of all the many particle systems, from macroscopic bodies like metal samples, to clusters, molecules, atoms and nuclei.

Collective oscillations determine to a large extent the cross-sections of interaction of all the above-mentioned objects with electromagnetic radiation and fast charged particles. For macroscopic bodies (solids, liquids and gases) they present the density oscillations, or acoustic waves, which are known best of all. The frequency of acoustic waves, $\omega_s$, is known to be proportional to the wave-vector, $k$.

It has long since been known, however, that in the macroscopic body of a subsystem, formed by light charged unbound particles with infinite range Coulombic repulsive interaction, the ordinary density or acoustic waves cannot exist. Instead, so-called plasma or Langmuir oscillations take place [1,2], with a completely different relationship between the frequency, $\omega_p$, and the wave-vector, $k$, from that of the frequency of acoustic waves. For small $k$ it has the following form [3]:

$$\omega_p^2 \approx \omega_{p0}^2 + \alpha k^2, \qquad (1)$$

where the so-called Langmuir or plasma frequency is presented by the following relation:

$$\omega_{p0}^2 = 4\pi e^2 n/\varepsilon m. \qquad (2)$$

---

In Eqs. (1) and (2) $\alpha$ is some constant, $e$ and $m$ are the value of the charge and the mass of the particle, $n$ is the number of the particles per unit volume and $\varepsilon$ is the dielectric constant of a conductor lattice (without the contribution of the conductive electrons). For metals it is usually assumed that $\varepsilon = 1$.

Langmuir oscillation is the simplest mode of a bulk uniform plasma oscillation in a conductor. It influences many physical properties of a sample, e.g., optical properties.

Plasma oscillations exist, as follows from the name, in plasmas and in electron gasses or liquids of metals and semiconductors [3, 4].

Eq. (1) was derived for a thin slab of a conductor [5] and for the infinite conductive media [6]. Hence it is valid for bulk samples of arbitrary shape when the wavelength, $\lambda$, is much shorter than the size of a sample, $a$ (1 « $ak$). In this case the infinite media approximation is reasonable. For long enough wavelengths the frequency, $\omega_p$, according to Eq. (1) is almost constant.

It is worthwhile to emphasize that collective frequencies, presented by Eqs. (1) and (2), do not include Planck's constant. Therefore the values of the frequencies can be understood in the frame of the classical approach (see, e.g. [6]). Quantum picture is important when we are interested in discrete quantum levels of the oscillations. The frequencies of the oscillation are the same in quantum and classical approach (compare results in [4] and [6]). But classical approach cannot give discrete quantum levels.

Two types of motion should be distinguished, the surface and the bulk oscillations. The surface oscillations are never accompanied by variation of the electron density, while some of the bulk oscillations are the oscillations of the density.

Mie had demonstrated long ago that in a bulk conductive sample of spherical shape the dipole oscillation of the collective electrons as a whole relative to the lattice of the ions does not exists [7]. Instead, the surface oscillation of a frequency √3 times smaller than $\omega_{p0}$ (Mie oscillation) exists in the bulk conductive samples of a spherical shape [7]. This occurs because oscillations with frequencies smaller than $\omega_{p0}$ cannot penetrate in the bulk of a conductive sample [5]. They decay in the surface layer of a sample [5]. The uniform (with wave-vector $k = 0$) dipole oscillation of the collective electrons as a whole relative to the lattice of the ions exists in thin slabs of conductive materials only [5] (see Fig. 1). In samples of more bulky shapes the restoring force acting on the collective electrons is not strong enough to provide frequency equal or larger than $\omega_{p0}$ [7].

Small elongation of the sphere leads to the splitting of a single Mie frequency into two, with a small difference between them [8].

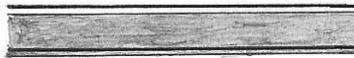

Fig. 1. Uniform dipole oscillation of the collective electrons as a whole relative to the lattice of the ions in the thin slab of a conductive material. Grey color shows collective electrons.

The problem of the uniform (with wave-vector $k = 0$) surface collective oscillation in essentially long shaped objects (nanotubes) was discussed in [9]. It was shown that the frequency of such oscillation decreases with the increase in the length of a nanotube. This frequency is inversely proportional to the square root of the length of a nanotube [9].

## 2. ELECTROSTATIC FIELD AND OSCILLATION IN A CONDUCTIVE ELLIPSOID

Let us consider electrically neutral bulk conductive ellipsoid with semi-axes $a$, $b$ and $c$. The volume of the ellipsoid $V = 4\pi abc/3$, the total charge of the ions is $enV$ (here $e > 0$ is the value of the electron charge, $n$ is the number of the collective electrons per unit volume of a sample). Electrostatic potential inside a uniformly charged ellipsoid (with total charge $enV$) can be found in [10]. For the ellipsoid of the ions of a considered sample we have:

$$\varphi_i(x,y,z) = enC - (2\pi en/\varepsilon)(N_a x^2 + N_b y^2 + N_c z^2), \tag{3}$$

where $C$ is some constant, $N_a$, $N_b$, and $N_c$ are the depolarization factors of the isotropic ellipsoid for the directions of the three axes [11].

As $N_a + N_b + N_c = 1$ [11], one can see that Eq. (3) satisfies the Poisson equation, $\Delta\varphi = -4\pi en/\varepsilon$.

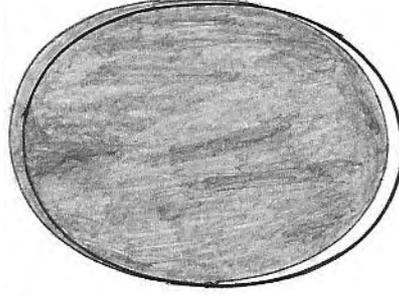

Fig. 2. Uniform dipole shift of the collective electrons as a whole relative to the lattice of the ions in the ellipsoid of a conductive material. Grey color shows collective electrons.

Now let us model the charge of the ions by a uniformly charged ellipsoid with the charge of the opposite sign with respect to the collective electrons charge sign (jellium model).

Let us calculate the electrostatic field in a sample when the ellipsoid of the collective electrons is shifted along the $x$-axis by small shift $-h$ relative to the ellipsoid of the ions. The electrostatic potential produced by the shifted ellipsoid of the collective electrons (of the same charge of an opposite sign) according to Eq. (3) is

$$\varphi_e(x,y,z) = -enC + (2\pi en/\varepsilon)[N_a(x + h)^2 + N_b y^2 + N_c z^2], \tag{4}$$

Total electrostatic potential is the sum of the potentials, produced by the ions and the collective electrons:

$$\varphi = \varphi_i + \varphi_e = (2\pi en N_a/\varepsilon)h(h + 2x). \tag{5}$$

The electrostatic field inside an ellipsoidal sample is equal to minus gradient of $\varphi$. It has $x$ component only:

$$E_i = -4\pi en N_a h/\varepsilon. \tag{6}$$

This field acts on each of the collective electrons with restoring force $-eE_i$.

Let us assume that the displacement of some current carrier is $h = h_0 \sin\omega_a t$. Then the Newton equation for the current carrier, $m\partial^2 h/\partial t^2 = -eE_i$, and Eq. (6) yield:

$$\omega_a = \omega_{p0} N_a^{1/2}. \tag{7}$$

It is well known [11] that for the ellipsoidal sample $N_a < 1$. Then, according to Eq. (7) the value of $\omega_a$ is smaller than that of $\omega_{p0}$. Hence such oscillation could not penetrate into the bulk of a sample. Regarded oscillation exists in the surface layer only (Mie oscillation).

When $a = b = c = R$ (spherical sample), $N_a = N_b = N_c = 1/3$ [11]. Eq. (7) accordingly yields $\sqrt{3}$ times smaller value of $\omega_a$ than that of $\omega_{p0}$ (Mie frequency). When $a$ is essentially smaller than $b$ and $c$, $N_a$ tends to 1 [11]. Eq. (7) in this case yields $\omega_a \approx \omega_{p0}$, like in the case of a thin slab. For $a$ essentially larger than $b = c$, $N_a$ tends to zero [11] along with the frequency $\omega_a$.

In general Eq. (7) describes 3 principal frequencies of the oscillations along the axes of a sample, $\omega_a$, $\omega_b$, and $\omega_c$. The frequencies $\omega_b$, and $\omega_c$ could be obtained from Eq. (7) by changing $a$ for $b$ and $c$. It is worthwhile to mention that as $N_a + N_b + N_c = 1$ [11],

$$\omega_a^2 + \omega_b^2 + \omega_c^2 = \omega_{p0}^2. \tag{8}$$

For spherical sample the principal frequencies are all equal, $\omega_a^2 = \omega_b^2 = \omega_c^2 = (1/3)\omega_{p0}^2$.

## 3. Discussion

Proposed approach allows calculating the frequencies of the Mie oscillations in a bulk sample of a shape of an arbitrary ellipsoid. It has been shown that the sum of the squares of the three principal Mie frequencies is equal to the square of the Langmuir frequency. The values of the principal Mie frequencies depend on the corresponding values of depolarization factors. It is well known that the depolarization factors $N_a$, $N_b$, and $N_c$ depend on the shape of a sample only and that they are expressed through certain elliptic integrals [11]. These integrals can be calculated for the ellipsoids of revolution (prolate and oblate spheroids) [11]. The depolarization factors of an arbitrary ellipsoid cannot be presented by elementary functions [11]. In [12] some simple approach was proposed in the frame of which some simple relation could approximate the depolarization factors:

$$N_a = b^2 c^2 / (a^2 b^2 + a^2 c^2 + b^2 c^2). \tag{9}$$

Two other depolarization factors can be obtained from Eq. (9) by cycling $a$, $b$ and $c$. As expected $N_a + N_b + N_c = 1$. When $a = b = c = R$ (spherical sample) Eq. (9) yields $N_a = N_b = N_c = N = 1/3$. When the eccentricity $\varepsilon$ is small Eq. (9) yields $N_a \approx [1 - (2/3)\varepsilon^2]/3$. This result is not essentially different from the expansion of the exact one, which yields $N_a \approx (1 - 0.4\varepsilon^2)/3$ [11]. When $a$ is essentially smaller than $b$ and $c$ Eq. (9) yields $N_a = 1$, like in the case of a thin slab. For $a$ essentially larger than $b = c$ $N_a \approx b^2/2a^2$, that is in this case the approximate value of $N_a$ is inversely proportional to the square of the length of a long axis, $a$.

# UNIFORMNE PLAZMENE OSCILACIJE U ELIPSOIDU PROVODNOG MATERIJALA

## Youri Kornyushin

*Razmatra se uticaj oblika uzorka na tip uniformnih dipolnih kolektivnih elektronskih oscilacija. U uzorcima "bulk" oblika uniformne dipolne oscilacije ne mogu da postoje, vec samo u uzorcima oblka tanke ploče. Ove oscilacije ne mogu da penetriraju elipsoidalne uzorke provodnih materijala, tako da se javljaju samo u površinskom omotaču uzorka ("Mie" oscilacije). Frekvence "Mie" oscilacija su izračunate za uzorke oblika proizvoljnog elipsoida.*